\documentclass[%
reprint,
 amsmath,amssymb,
showkeys,
longbibliography
]{revtex4-1}
\usepackage{graphicx}
\usepackage{dcolumn}
\usepackage{bm}
\usepackage{siunitx}
\usepackage[final]{pdfpages}
\usepackage{pgffor}
\setboolean{@twoside}{false}

\usepackage{natbib}
\makeatletter
\AtBeginDocument{\let\LS@rot\@undefined}
\makeatother

\begin{document}

\title{Flow rate of transport network controls uniform metabolite supply to tissue}

\author{Felix J. Meigel}
\author{Karen Alim}%
 \email{karen.alim@ds.mpg.de}
\affiliation{%
 Max Planck Institute for Dynamics and Self-Organization, D-37077 G\"ottingen, Germany
}%

\date{\today}

\begin{abstract}
Life and functioning of higher organisms depends on the continuous supply of metabolites to tissues and organs. What are the requirements on the transport network pervading a tissue to provide a uniform supply of nutrients, minerals, or hormones? To theoretically answer this question, we present an analytical scaling argument and numerical simulations on how flow dynamics and network architecture control active spread and uniform supply of metabolites by studying the example of xylem vessels in plants. We identify the fluid inflow rate as the key factor for uniform supply. While at low inflow rates metabolites are already exhausted close to flow inlets, too high inflow flushes metabolites through the network and deprives tissue close to inlets of supply. In between these two regimes, there exists an optimal inflow rate that yields a uniform supply of metabolites. We determine this optimal inflow analytically in quantitative agreement with numerical results. Optimizing network architecture by reducing the supply variance over all network tubes, we identify patterns of tube dilation or contraction that compensate sub-optimal supply for the case of too low or too high inflow rate. 
\end{abstract}

\keywords{Biological Physics, Fluid Dynamics}
\maketitle
Transport processes organized in networks structures are ubiquitous in our life, from road traffic \cite{sen_small-world_2003} and power networks \cite{rohden_self-organized_2012} to river estuaries \cite{rodriguez-iturbe_fractal_1997} and vascular systems of extended organisms \cite{netti_macro-_1997,kapellos_chapter_2015}. Especially fluid flow driven transport through networks is underlying many technological applications like fuel cells \cite{MA2005418}, micro-fluidic devices \cite{santini_controlled-release_1999} or filtration systems \cite{tufenkji_correlation_2004} and their medical applications  \cite{levey_new_2009}. Most significantly, all higher forms of life rely on a fluid flow based transport networks to provide their tissue with metabolites like nutrients or minerals, as there are the circulatory system of animals \cite{isogai_vascular_2001}, the plant xylem vascular system \cite{PlantsinAction:chap3}, and the hyphae networks of fungi \cite{boddy_saprotrophic_2009,tero_rules_2010, heaton_advection_2012}. Within a tissue, each cell needs to be provided with the same minimal amount of metabolites. How does a transport network need to be set up to make sure that metabolites arrive uniformly at each cell within a tissue? Here, we theoretically investigate the requirements on flow and network architecture for uniform supply.

On the level of inter-vascular tissue, models for minimal supply due to metabolites uptake and metabolite diffusion within the tissue date back one hundred years to Krogh's model \cite{krogh_number_1919}. Yet, Krogh assumes that metabolites are provided by the vasculature at a constant rate at all vessel walls \cite{Rubenstein:2012}. This strong simplification neglects that vascular network architecture and resulting asymmetries in flow based transport give rise to large variations in metabolite availability within the network. On the level of the vascular network itself, studies mapping out variations in metabolite availability are scarce \cite{fang_oxygen_2008, schneider_tissue_2012, marbach_pruning_2016}. Insight, what controls uniform metabolite supply at the vasculature level is missing. Instead research focused on network flow and not transport properties identifying scaling relationships regarding the network's fluid dynamics \cite{sack_leaf_2013,sack_hydraulic_2004,coomes_scaling_2008,price_scaling_2012,jensen_physical_2013,labarbera_principles_1990,west_general_1999,murray_physiological_1926}. Another branch of theoretical models for vascular systems investigated optimal network topologies with minimal transport cost in the form of dissipation \cite{durand_architecture_2006,west_general_1999,bohn_structure_2007}. Including robustness to damage or flow fluctuations \cite{katifori_damage_2010,corson_fluctuations_2010} or vessel growth \cite{ronellenfitsch_global_2016} in these studies resulted in network topologies in closer resemblance to observable vascular networks. However, despite the efforts to find minimal dissipation networks to understand transport networks in nature, it is not obvious that the efficiency of a transport network is what organisms optimize for. Instead measurements on Zebrafish vasculature suggest that biological transport networks care for uniform partitioning of blood cells and thus uniform oxygen supply \cite{chang_optimal_2015} and are not optimized for minimal dissipation. The vasculature architecture might, as established for tissue, be build for the uniform supply of metabolites like oxygen, nutrient compounds or biochemical signals. A further indication is that networks adapt to reinforce supply to tissues and organs \cite{Chan2002257}.

The spread of metabolites through fluid flow is well-described in hydrodynamics. Here, the transport of particles in a single long slender tube is efficiently captured by \textit{Taylor Dispersion} \cite{taylor_dispersion_1953,Aris:1956,bruus_theoretical_2009}. The important contributions of particle transport are advection by fluid flow and molecular diffusion, resulting in a decoupling of flow dynamics and particle concentration dynamics. Transport of particles that in addition get absorbed along the tube wall is well-studied in the setting of heat conduction \cite{e._m._lungu_effect_1982,graetz_ueber_1882,HeatAbsoChem,hemida_theoretical_2002}. 
Yet, the concentration patterns of particles within a transport network is fundamentally more complicated due to the particle concentration being coupled in a global manner by the network spanning flow. Thus, further theoretical development is required as presented here.

\begin{figure}
\centering
\includegraphics[width=\linewidth]{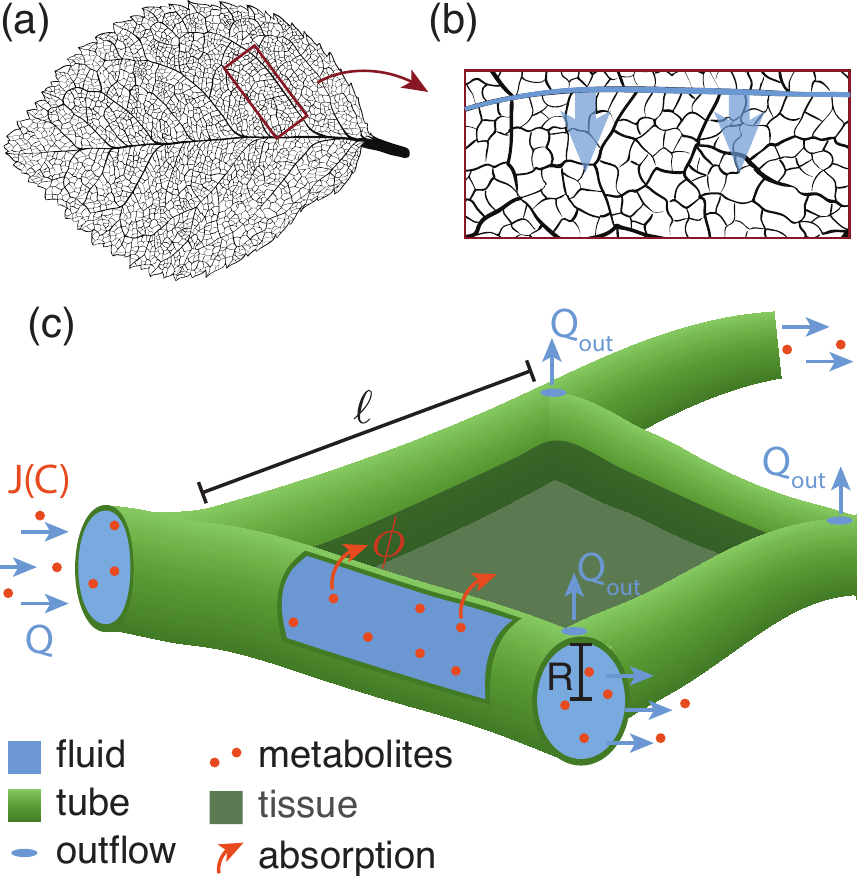}
\caption{Schematic sketch of metabolite supply in leaves. (a) Vasculature of a leaf displaying the primary vein horizontally at the center and secondary veins as next biggest veins departing from the primary vein, down to the highly inter-webbed higher order veins. (b) The secondary vein (\textit{blue}) supplies the tubular higher order vein network with metabolites and fluid. (c) Xylem vessel network modeled as network of tubes of varying radius $R$ and length $\ell$. Inflow of fluid flow rate $Q$ and metabolite flux $J$ from upstream tubes (\textit{left}). Fluid evaporation through stomata a the leaf surface modeled by constant outflow $Q_{out}$ at every network node. Metabolites are advected and diffuse within the fluid. In addition, metabolites get absorbed $\phi$ along the tube wall into cells at a constant rate $\nu$.}
\label{Fig:cartoon}
\end{figure}

While a hydrodynamic perspective provides a general picture with a minimum of assumptions and hence a broad applicability, the specifics of metabolite flow differs between biological systems. In the light of network optimization approaches \cite{katifori_damage_2010,corson_fluctuations_2010,ronellenfitsch_global_2016}, we apply the general hydrodynamic perspective to the tissue specifics of plant leaf xylem vessels in dicotyledons. In plants, the transport of water and metabolites, especially soil bound nutrients and minerals like nitrate or potassium \cite{PlantsinAction:chap3}, from the plant roots to the leaf tissue is routed in highly pitted and rigidly lignified xylem veins \cite{zwieniecki_hydraulic_2002}. Xylem veins should not to be confused with the oppositely routed phloem veins predominantly transporting sugar away from the leaf tissue \cite{roth-nebelsick_evolution_2001,jensen_osmotically_2009}. We consider the spread of the scarce metabolites in xylem fluid as limiting for maintaining the function of leaf cells and thus focus on the xylem network neglecting the detailed spreading dynamics of metabolites within the tissue itself. Metabolites enter leaf cells dominantly at the level of inter-webbed higher order veins, while primary and secondary order veins distribute metabolites over the large scale of the leaf \cite{holbrook_vascular_2011}.
Here we focus on a leaf tissue excerpt pervaded by higher order veins.
A secondary vein is the source of metabolite enriched fluid flowing through higher order vessels pervading the leaf tissue, see Fig.~\ref{Fig:cartoon} (a), (b). Fluid flow is regulated through evaporation control across the entire leaf blade  \cite{fricker_stomata_2012}. 
Evaporation is commonly modeled by a constant outflow of fluid at every node within the vascular network \cite{katifori_damage_2010,corson_fluctuations_2010}, for details see supplementary material S1.  Metabolites are absorbed continuously along the walls of the tubular vessels into the tissue supplying the cells there.  

In this paper, we develop a theoretical framework to optimize flow dynamics and architecture of transport networks for uniform supply of metabolites to surrounding tissue focusing exemplarily on plant xylem networks. We derive analytical expressions for the absorption of metabolites within a single tube  and use these results to simulate supply patterns in the inter-webbed transport networks. We find that the inflow rate is the dominant factor controlling supply patterns for all network architectures. For low inflow rate, average fluid velocities are small and metabolites are mainly absorbed next to the flow inlets. For high inflow rates, average velocities are fast and metabolites are mainly absorbed at the far end opposing the inlets. In between, we identify an optimal inflow rate, that yields uniform absorption and thus supply levels. We present a one-dimensional network analogue that allows us to derive an analytical expression for the optimal flow rate as a function of system parameters such as network size and average tube radius in agreement with simulations. Further optimizing the network architecture for low, optimal, or high inflow rate, we find that localized adaptation in tube radius is capable to compensate for the non-uniform supply patterns at low and high inflow rate, yet cannot outcompete the optimal inflow rate.
\section*{Results}
\subsection*{Metabolite absorption across a fluid filled tube}

\begin{table}
\caption{Nomenclature}
\begin{tabular}{|l l|}
\hline
$R$ & tube radius\\
$\ell$ & tube length\\
$L$ & network length\\
$N$ & number of tubes in the network\\
$Q$ & fluid flow rate\\
$U$ & fluid flow velocity\\
$C$ & metabolite concentration\\
$J$ & metabolite flux \\
$\kappa$ & molecular diffusivity of metabolite\\
$\nu$ & absorption rate of metabolite\\
$\gamma$ & absorption parameter; $\gamma=\nu/\kappa$ \\
$\phi$ & overall absorption along a tube \\
$\hat{\phi}$ & overall absorption capacity along a tube\\
Pe & P\'eclet number, ratio of diffusive and advective time scale\\
$S$& ratio of the time scales for absorption\\
$\beta$ & concentration decay constant\\
\hline
\end{tabular}
\end{table}

Consider a cylindrical tube filled with fluid flowing at flow velocity $U(r,z)$ along the tube. Metabolites are advected with the flow and in addition disperse due to molecular diffusion with diffusivity $\kappa$.  
Considering the small scales of xylem vessels and xylem flow, we are in the regime of low Reynolds number, where flow is best described by laminar Poiseuille flow. Thus, we describe the flow along the longitudinal axis $z$ varying in radial direction $r$ in the circular tube with a radius $R$ and length $\ell$ as, $U(r,z)=2 \left(1- \frac{r^2}{R^2} \right) \langle U(z) \rangle_r $. Here, $\langle U \rangle_r$ denotes the cross-sectionally averaged longitudinal velocity. The spread of metabolites of concentration $C$ is thus fully described by
\begin{align}
\frac{\partial C}{\partial t} + U(r,z) \frac{\partial C}{\partial z} = \kappa \left[\frac{1}{r} \frac{\partial}{\partial r}\left(r \frac{\partial C}{\partial r}\right)+\frac{\partial^2 C}{\partial z^2} \right].
\label{eq:Absorb:Dispersion}
\end{align}
Metabolites are absorbed into the surrounding tissue along the tube wall, given by the boundary condition
\begin{align}
\kappa \frac{\partial C}{\partial r}\bigg|_{r=R} + \nu C(R) = 0, \label{eq:Absorb:Boundary}
\end{align}
analogous to heat absorption or surface reactions. Here, the parameter $\nu$ denotes the metabolite absorption rate at the tube wall. Dividing $\nu$ by the molecular diffusivity we define the absorption parameter $\gamma$, where we consider $\nu$ as a constant tissue property. According to \cite{e._m._lungu_effect_1982}  the advection-diffusion equation, Eq.~\eqref{eq:Absorb:Dispersion}, can be re-formulated employing both the boundary condition Eq.~\eqref{eq:Absorb:Boundary} and the Poiseuille profile as a single absorption-advection-diffusion equation in cylindrical coordinates. 

In analogy to the derivation of \textit{Taylor Dispersion} by G.I.~Taylor \cite{taylor_dispersion_1953}, a simpler, though approximated, expression is possible where the concentration dynamics only depend on the longitudinal coordinate $z$. To this end the metabolite concentration is separated into the sum of a cross-sectional average concentration $\langle C \rangle_r$ and the radial variation $C'$, $C(r,z)~=~\langle~C(z)~\rangle_r~+~C'(r,z)$. The multidimensional diffusion-advection for $C(r,z)~=~\langle~C(z)~\rangle_r~+~C'(r,z)$ can be simplified to an equation for the cross-sectionally averaged concentration $\langle~C(z)~\rangle_r$ if the cross-sectional variations of the concentration $C'(r,z)$ are much
smaller than the averaged concentration itself \cite{taylor_dispersion_1953, Cha_inprep}, resulting in
\begin{align}
\frac{\partial \langle C \rangle_r}{\partial t} =& - \frac{2 \kappa}{R^2} \frac{4 \gamma R}{ 4+ \gamma R} \langle C \rangle_r  - \frac{12 + \gamma R}{12 + 3 \gamma R} \langle U \rangle_r  \frac{\partial\langle C \rangle_r}{\partial z} \nonumber \\ 
&+ \left( \kappa + \frac{12 + \gamma R}{12 + 3 \gamma  R} \frac{\langle U \rangle_r^2 R^2 }{48 \kappa} \right) \frac{\partial^2 \langle C \rangle_r}{ \partial z^2}.  \label{eq:absorb:effaddifeq} 
\end{align}
This approach employs three approximations. First, the time-scale of diffusion across the tube's cross-section has to be much smaller than the time-scale of advection, $\frac{\ell}{\langle U \rangle_r} \gg \frac{R^2}{\kappa}$. This sets an upper bond for the later choice of the fluid inflow rate. 
The second assumption states that the cross-sectional variations of the concentration have to be small $\langle C \rangle_r \gg C'$. Since a high absorption parameter $\gamma$ implies a large concentration gradient across the cross-section, the second assumption implies  $\gamma R \ll 1$. 
Third, the variation of $C'$ has to be much greater in radial direction than in flow direction $ \partial^2_r C' \gg \partial^2_z C'$. The last assumption implies the tube radius to be smaller than its length $R \ll \ell$. This is fulfilled by long slender tubes.

Employing these assumptions, the cross-sectional average metabolite concentration along a tube in steady state is given by an exponential decay from initial concentration $C_0$,
\begin{align}
\langle C(z) \rangle_r &= C_0 \exp(- \beta \frac{z}{\ell} ), \label{eq:Absorb:steadysolution} \\ 
\beta &=  \frac{24 \cdot \text{Pe} }{48  +  R \gamma \cdot \frac{\text{Pe}}{S} } \left( \sqrt{1+ 8 \cdot \frac{S}{\text{Pe}}+ \frac{3}{4} \cdot R \gamma} -1 \right). \nonumber
\end{align}
Here, we introduced two dimensionless variables $\text{Pe}$ and $S$. $\text{Pe}= \frac{\langle U \rangle_r \ell}{\kappa}$ is the well known Peclet number describing the relation between diffusive and advective time scale.
$S$ is the ratio of the time scale for absorption, given by the product of dimensionless absorption parameter and time to diffuse across the tube's cross-section, and the time to be advected out of the tube, resulting in $S=\frac{\gamma \kappa \ell}{R \langle U \rangle_r}$.

Considering a constant influx $J_0$ by advection and diffusion of metabolites at the tube's start, we find that the initial concentration $C_0$ is given by
\begin{align}
C_0= \frac{J_0}{\langle U \rangle_r + \kappa \frac{\beta}{\ell}}.
\label{eq:Absorb:Initial}
\end{align}

The overall absorption $\phi$ along a tube is given by the integrated flux of metabolites across the tube wall $\mathcal{W}$, $\phi =2 \pi R \int_{\mathcal{W}} \kappa \nabla C \text{d}z$, where $\text{d}z$ is integrating over the length of the tube. As in the derivation of the effective diffusion-advection-absorption equation Eq.~\eqref{eq:absorb:effaddifeq}, we use $R \gamma \ll 1$ to arrive at
\begin{align}
\phi =& \pi R^2 J_0 4 \frac{S}{\text{Pe}} \left(48 + \frac{R \gamma \cdot \text{Pe}}{S} \right) \nonumber \\
&\cdot \frac{\left(48 - \frac{R \gamma \cdot \text{Pe}}{S} \left( \Lambda -2 \right) \right)}{48 (48+\frac{\gamma R \cdot \text{Pe}}{ S}+24 (\Lambda-1)) \cdot (\Lambda-1)} \nonumber \\ \cdot&   \left(1 - \exp \left( -24 \cdot \text{Pe} \cdot \frac{\Lambda-1}{48 + \gamma R \text{Pe} / S} \right)  \right), 
\label{eq:Absorb:Grand_Absorb}
\end{align}
where $\Lambda$ is an abbreviation $\Lambda=\sqrt{8 S/\text{Pe}+3/4 \gamma R+1}$. 
We identify two factors that control the absorption in a tube. The first is the total influx of metabolites over the cross-sectional area of the tube $\pi R^2 J_0$. The total influx of metabolites is the upper limit for absorption in the tube. The second factor is the tube's capacity to absorb metabolites as $\hat{\phi} = \phi/\pi R^2 J_0$ with $\hat{\phi} \in [0,1]$. This absorption capacity is independent of the concentration of metabolites and only depends on the parameters of the tube and the flow velocity within the tube.

For the derivation of the optimal inflow rate, it is essential to approximate the absorption capacity as resulting from Eq.~\eqref{eq:Absorb:Grand_Absorb} above. We initially approximate the inverse of the absorption capacity $\hat{\phi}^{-1}$ by taking a finite $\text{Pe}>0$ and using $R \gamma \ll 1$ to find $\hat{\phi}^{-1}=\frac{1}{2 S} +1$. Resubstituting the system's parameters for $S$ we find for the absorption capacity of a tube
\begin{align}
\hat{\phi}=\frac{2 \gamma \kappa \ell}{R \langle U \rangle_r +2 \gamma \kappa \ell}.
\label{eq:Absorb:simplified}
\end{align}
Note the simple dependence of the absorption capacity on the cross-sectionally average flow velocity in the tube. The approximation of the absorption capacity has been verified numerically to hold over the parameter space considered here, see supplemental information S4. Note, that this simplified expression is only used for the analytical derivation of the optimal inflow rate. For simulations the full expression Eq.~\ref{eq:Absorb:Grand_Absorb} is used. From now on, we drop the brackets $\langle \rangle_r$ and only refer to cross-sectional averaged observables.   
\subsection*{Absorption patterns in fluid flow driven transport networks}
\label{sec:Supply_Pattern_Uniform}

\begin{figure}[!tpbh]
    \includegraphics[width=\linewidth]{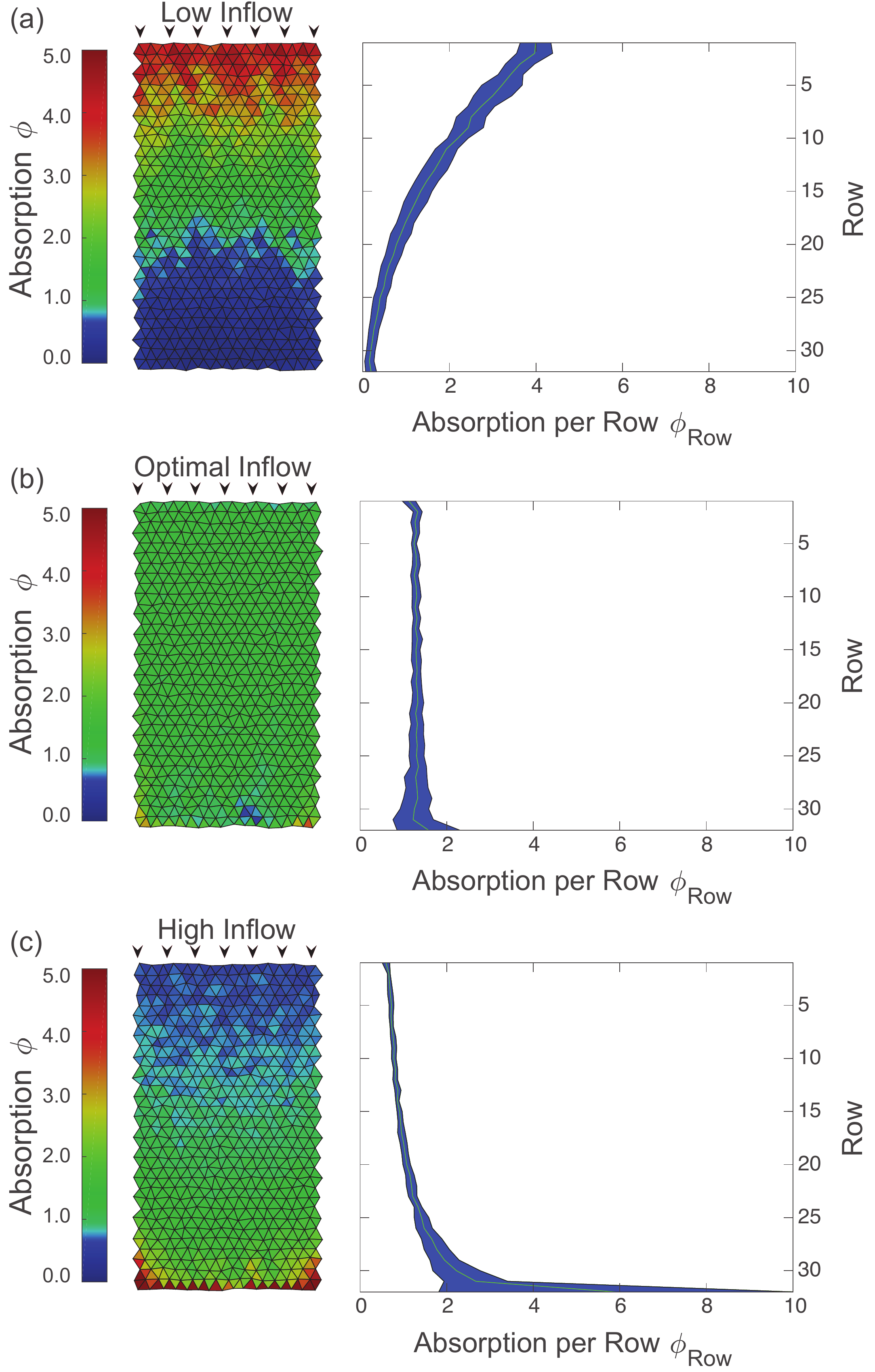}
    \caption{Supply patterns are controlled by fluid inflow rates. Supply pattern of a rectangular tissue section pervaded by a transport network for increasing fluid inflow rate ranging from (a) $Q_{\text{in}}=\SI{0.8e-6}{\milli\metre\cubed\per\second}$, via (b) $Q_{\text{in}}=\SI{3.2e-6}{\milli\meter\cubed\per\second}$, to (c) $Q_{\text{in}}=\SI{6.4e-6}{\milli\meter\cubed\per\second}$. The transport network is build of tubes of equal radius and roughly equal length triangulating the tissue section under consideration. Metabolites are absorbed across tube walls into the tissue. Left column: Supply pattern in every triangulated tissue section given by the average metabolite absorption along neighboring tubes. The absorption is normalized with the inverse of the total influx $J_{\text{tot}}^{-1}$ and the total number of tubes $\mathcal{N}$. Right column: Standard deviation and mean absorption per row counting downward from the inflow nodes at the top of the network.  At low inflow rate (a) metabolites are absorbed close to inflow and are not transported through the network while for high inflow rate (c) metabolites get flushed through the network for being absorbed mainly at the end. The variance in absorption across all tubes is $0.75$ for low inflow rate and $2.35$ for high inflow rate. In between these two cases an optimal inflow rate with the lowest variance exists (b) that yields uniform supply and a overall variance of only $0.07$. Remaining metabolites are flowing out at the bottom end amounting to $0.6\%$, $4.2\%$, and $19.4\%$ of the metabolite influx for (a), (b), and (c), respectively.} \label{Fig:Uniform}
\end{figure}
  In a transport network, individual tubes are connected at nodes. Here, we aim to model the geometry of higher order xylem veins branching from a second order vein in dicotyledons, as shown in Fig.~\ref{Fig:cartoon} b). We choose a planar transport network representing a rectangular excerpt of the leaf tissue. For the tissue excerpt we choose a general vascularisation using a slightly randomized tessellation of space, where the network is build with small triangles, known to bear least artifacts \cite{katifori_damage_2010}. A small Gaussian noise of a twentieth tube length $\ell$ is added to the positions of the tessellation nodes to avoid pattern artifacts arising from the underlying topology otherwise, see supplementary material S2. The tube length varies accordingly in a normal distribution around the mean tube length $\langle\ell\rangle$. In agreement with observations of diminishing hierarchy in the higher interwebbed xylem vessel radii \cite{ronellenfitsch_topological_2015,sack_leaf_2013}, we set the same radius $R$ for all tubes. Fluid and metabolites are flowing into the network at network nodes along one side of the rectangular region, representing the supply from secondary veins into the tissue. Xylem vessels are organized in vascular bundles in lower order veins that branch out into the interwebbed higher order xylem network \cite{thorne_structure_2006} presumingly supplying the same flow at every inflow node. Therefore, we approximate inflow rates $Q_{\text{in}}$ to be equal at all inflow nodes. To represent the effect of fluid evaporation at stomata, fluid, but not metabolite, is flowing out at every node in the network $Q_{\text{out}}$, see Fig.~\ref{Fig:cartoon}~c). Metabolites are absorbed across each tube wall. The absorption rate $\nu$ is constant throughout the network. Yet, as we have learned by studying a single tube in the previous section, the amount of metabolite absorbed depends on how much metabolite available in the fluid, and how much time the metabolite has to travel to the tube wall to get absorbed. Therefore, absorption despite a constant absorption rate varies largely within a network.

The flow of the metabolites is determined by the fluid flow in accordance with Eq.~\ref{eq:Absorb:Dispersion}. The fluid flow throughout a network is fully defined by the network's architecture, the inflow and outflow rates and Kirchhoff's circuit law. The cross-sectionally averaged fluid flow velocity in a tube follows subsequently from pressure drops $\Delta P= R_{\text{hyd}} \cdot Q$ along the tube. 
Each tube is considered as straight cylinder with hydraulic conductance of $K_{\text{hyd}}^{-1} = \frac{8}{\pi} \ell \eta \frac{1}{R^4}$, where $\eta$ denotes the dynamic fluid viscosity. The pressure at every node is calculated by multiplying the inverse of the network conductivity matrix with the inflow or outflow rates at every node. The pressure drop along the tube is the difference between the pressure values at start and end node. The fluid flow is solved consistently through the whole network and takes network geometry and viscous and friction forces via the hydraulic resistance into account. Considering steady state solutions, the flow does not fluctuate over time.

The absorption of metabolite across a tube's wall within the network depends on the metabolite available and the tube's absorption capacity. The absorption capacity of tube follows directly from each tubes physical parameters and the fluid flow velocity within the tube. Next, we need to calculate the influx of metabolites $J_0$ in every tube which we solve for iteratively throughout the network starting with the influx nodes, see supplementary material S3A. As simplification we focus on stationary, steady state absorption patterns.  We employ that metabolite flux is conserved at every network node. All metabolites flowing into an node are redistributed into tubes originating from this node. Redistribution is proportional to diffusion and flux into each tube. The metabolite outflux at the end of a tube is then given by the difference of metabolite influx and total absorption along the tube. Finally, at the lower end of the considered network excerpt, opposite the inflow nodes, remaining metabolites are flowing out of the network. Since the outflowing metabolites would lead to an accumulation of metabolites, we state the amount of metabolites not absorbed for every considered network excerpt.

Taking our initial motivation from plant leaves, we choose an average tube length $\ell=\SI{0.1}{\milli\meter}$ and tube radius $R=\SI{3}{\micro\meter}$ in accordance with xylem vessels \cite{sack_leaf_2013,sack_hydraulic_2004,sack_developmentally_2012}. Note that there is a difference between leaf vein and xylem vessel radius, as leaf veins bundle both phloem and xylem vessels. We thus consider xylem vessels to be less than half the radius of leaf veins in our parameter choice. The order of magnitude of the total inflow rate is chosen to yield velocities observable in lower order xylem vessels $\langle U\rangle_r \approx \SI{1}{\micro\meter\per\second}$. We vary the fluid inflow rate between $Q_{\text{in}}~=~\SI{0.8e-6}{\milli\meter\cubed\per\second}$ to $Q_{\text{in}}~=~\SI{6.4e-6}{\milli\meter\cubed\per\second}$. The choice of the inflow is consistent with a average water evaporation of approximately $\SI{0.1}{\mol\per\meter\squared\per\second}$ \cite{vico_effects_2011,zwieniecki_hydraulic_2002} and an average stomata density of~$\SI{200}{\per \milli \meter \squared}$ \cite{woodward_influence_1995}.   
For the molecular diffusivity, we consider small molecules with $\kappa=\SI{1e-10}{\meter\squared\per\second}$. For the network size, we choose a triangulation with $\mathcal{N} \approx 1000$ tubes.
Since xylem vessels consist of highly pitted dead lignified tissue, no active absorption by chemical reactions but passive absorption by membrane permeation is expected. Values for membrane permeation are typically in the range of $\nu \approx \SI{1e-9}{\meter}$ \cite{phillips_physical_2013} and depend on both membrane and metabolite properties.
Alternatively an estimation for the absorption parameter $\gamma$ can be derived from concentration profiles in xylem veins \cite{vanBel:1979}. Translating the measured exponential decay for higher order veins, we find $\gamma \approx \SI{10}{\per\meter}$ \cite{Horwitz:1958, vanBel:1979}. This estimate for $\gamma$ is in accordance with estimates using the membrane permeation. Thus, for the numerical calculations an absorption parameter of $\gamma =\SI{10}{\per \meter}$ is chosen. Revisiting the three assumption made in Eq.~\ref{eq:absorb:effaddifeq} we verify that these assumptions hold for the chosen network topologies, see supplementary material S4.

\begin{figure*}[!tbh]
\centering
\includegraphics[width=1\linewidth]{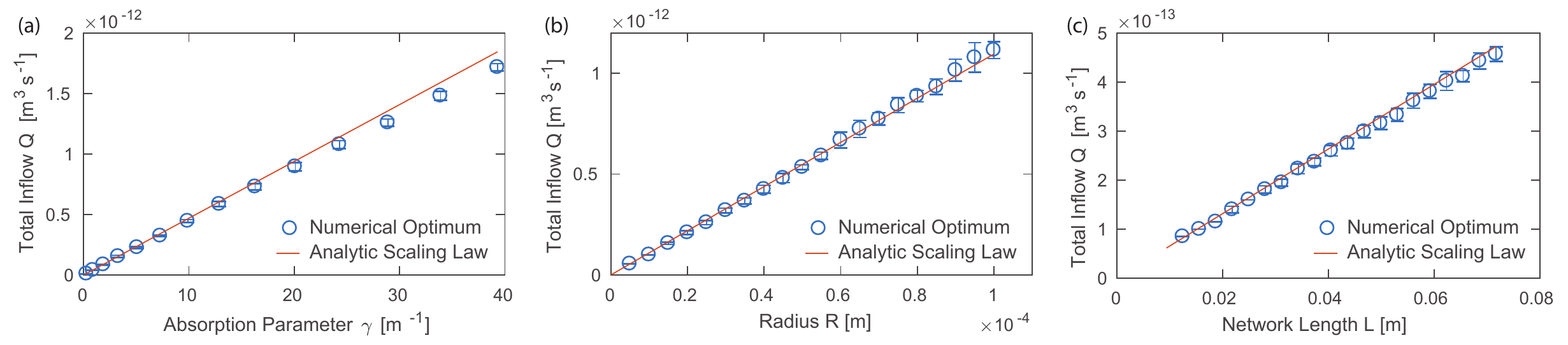}
\caption{Scaling of the optimal inflow rate for uniform supply of a two-dimensional rectangular tissue section with system parameters. Optimal inflow rate defined as lowest overall variance in absorption scales linearly with the absorption parameter $\gamma$ (a), the total network length $L$ (b) and the tube radius $R$ (c). While one parameter was varied, the other parameters were kept constant. Error bars represent the standard deviation over 15 independent runs. Data is in agreement with the scaling law (\textit{red line}) of optimal inflow rate derived for a one-dimensional toy model and adapted to two-dimensions by a geometrical factor. 
}
\label{FIG:Leaky}
\end{figure*}
We first study metabolite supply patterns in uniform transport networks, where all tubes have the same radius $R$. To compare different inflow rates we normalize the absorption by the total influx of metabolites. We find that the total fluid inflow rate is dominating the supply patterns, see Fig.~\ref{Fig:Uniform}. For small inflow rate, average flow velocities in the network are slow and the highest absorption is near the inflow nodes. Metabolites are not transported through to the end of the network limiting supply there. Calculating the mean absorption per row $\phi_{\text{row}}$ from inflow to opposing end, we can characterize this regime by $\phi_{\text{row}} >\phi_{\text{row}+1}$, see Fig.~\ref{Fig:Uniform} right column. For high inflow rates, average flow velocities are fast and the absorption increases with the distance from the inflow nodes $\phi_{\text{row}} <\phi_{\text{row}+1}$. Metabolites arrive at the end of the network too quickly before getting absorbed limiting supply close to the inflow nodes. Between these limiting cases, we identify an inflow rate that gives rise to an optimally uniform supply pattern. We define the optimum by the lowest variance. The variance is calculated over the ensemble of all tubes. The overall variance in the optimal case is $0.07$ compared to $2.35$ and $0.75$ in the examples of low and high inflow rate shown in Fig.~\ref{Fig:Uniform}. In the optimally uniform supply pattern, absorption is the same constant rate in subsequent rows, $\phi_{\text{row}} =\phi_{\text{row}+1}$. On the basis of this simple relation stating uniform absorption, a scaling law for the optimal inflow rate is derived next.
\subsection*{Scaling law for the optimal inflow rate}
To derive a scaling law for the optimal inflow rate that gives rise to the most uniform supply pattern, we consider a one-dimensional toy model of connected single tubes that captures the essential flow and transport characteristics along the rows of the two-dimensional transport networks investigated above. For this, we look at a straight pipeline of $N$ identical tubes. As for the networks considered in the previous section, see Fig.~\ref{Fig:Uniform}, all tubes are of the same radius $R$ and length $\ell$, in accordance with observations of diminishing radius hierarchy in higher order veins \cite{ronellenfitsch_topological_2015,sack_leaf_2013}. Metabolites and fluid are flowing into the first tube $Q_{\text{in}}$ and fluid is leaving at a constant rate $Q_{\text{out}}$ at every node between adjacent tubes. Metabolites cannot exit at nodes but remain in the fluid until the very end of the pipeline or are absorbed. Also the fluid inflow rate and total fluid outflow rate are equal, i.e.~$Q_{\text{out}}=Q_{\text{in}}/N$. This results in a constant decrease in flow rate by $Q_{\text{in}}/N$ from one tube to the subsequent. To translate this to cross-sectionally averaged flow velocities, which are the flow properties determining absorption, we use $U=Q/\pi R^2$. Consequently, the flow velocity in segment $m+1$ is given by $U_{m+1}=U_m - Q_{\text{in}}/\pi R^2N$. 

The outflux of metabolites from one tube is equal to the influx of metabolites in the subsequent tube $J_{\text{out},m} = J_{\text{in},m+1}$, since all tubes have the same radius $R$.
If $\pi R^2 J_0$ is the total amount of metabolites flowing into the first tube, then only the fraction $1-\hat{\phi}_1$ is flowing out while the fraction $\hat{\phi}_1$ is absorbed. Generalizing we determine the absorption in tube $m$ as
\begin{align}
\phi_m= \pi R^2 J_0 \hat{\phi}_m \prod_{j=1}^{m-1} (1-\hat{\phi}_j).
\label{eq:Absorb:one-dim}
\end{align}
The state of optimally uniform absorption is now defined by absorption in subsequent tubes being equal. We use this constraint to determine the inflow rate that corresponds to the optimally uniform supply pattern. Using Eq.~\eqref{eq:Absorb:one-dim} to write the absorption in the $m+1$ tube as a function of the absorption in the previous tube and inserting the equality constraint, we arrive at an expression including absorption capacities only, $(1-\hat{\phi}_m) \hat{\phi}_{m+1}=\hat{\phi}_m$. Inserting the simplified expression for the absorption capacity from Eq.~\eqref{eq:Absorb:simplified}, we find the scaling law determining the optimal inflow rate to yield uniform absorption
\begin{align}
Q_{\text{in}}=2\pi\kappa \gamma L R, \label{eq:ScalingLaw}
\end{align}
where $L=N\ell$ denotes the length of the sequence of pipes. Note, that this condition is independent of which tube segment $m$ is considered. The absorption is uniform along the entire sequence of tubes. 

\begin{figure*}[!htb]
  \begin{minipage}[c]{0.60\textwidth}
    \includegraphics[width=\textwidth]{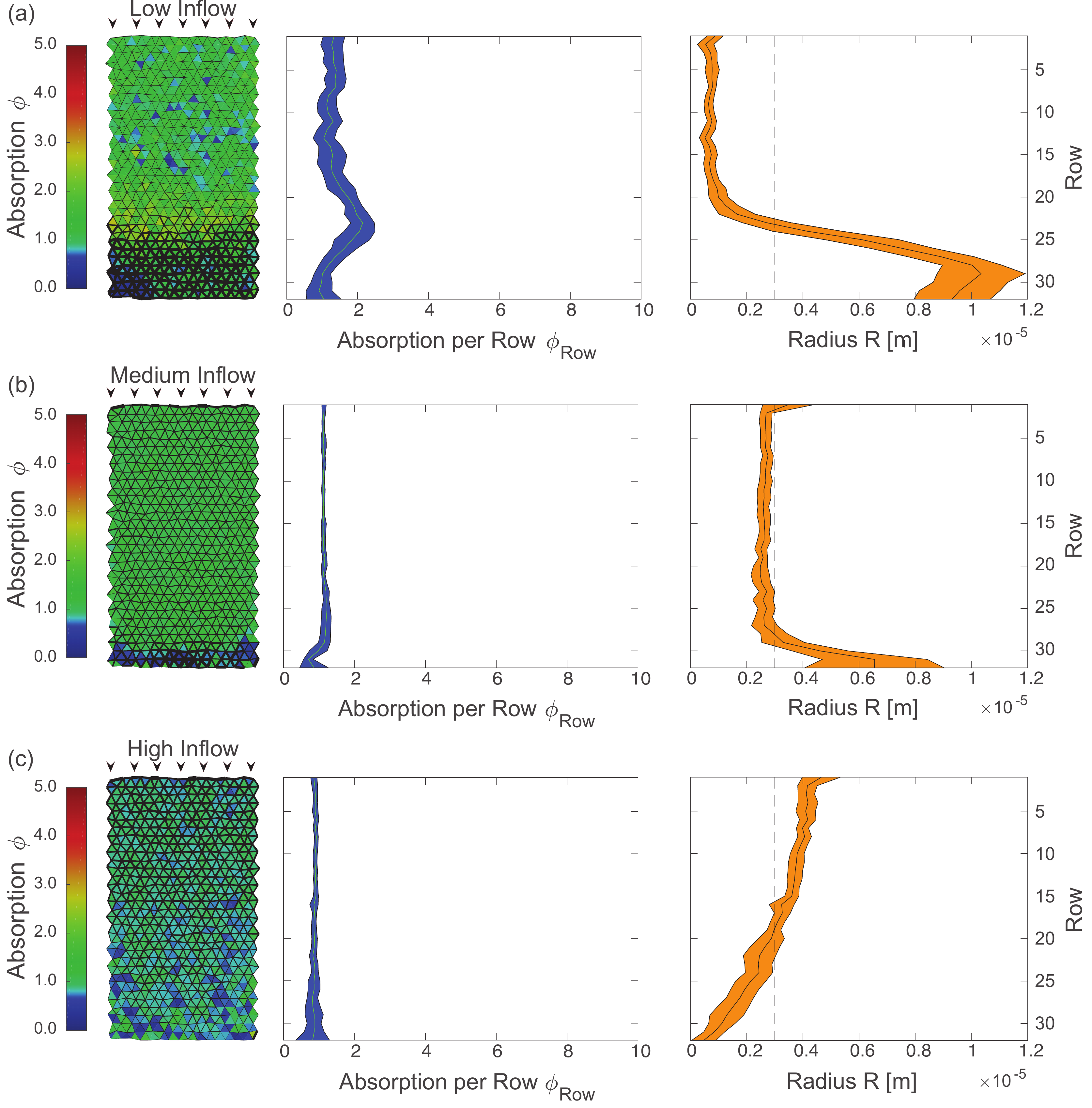}
  \end{minipage}\hfill
  \begin{minipage}[c]{0.37\textwidth}
 \caption{Optimized network architectures for uniform metabolite supply patterns. Supply pattern for the same low (a), optimal (b) and high (c) inflow rate as in Fig.~\ref{Fig:Uniform} but optimized network architecture. Left column: Supply pattern in every triangulated tissue section given by the average metabolite absorption along neighboring tubes, see also Fig.~\ref{Fig:Uniform}. Thickness of tubes represents the tube radius. Middle column: Standard deviation and mean absorption per row counting downward from the inflow nodes at the top of the network. Right column: Standard deviation and mean radius per row. Dashed line marks average tube radius. (a) For low inflow rate tubes contract near inflow nodes, speeding up flows there and thus propagating metabolites further down the network. Tubes dilate toward the network end further increasing absorption there. Variance in absorption is reduced to $0.211$. (c) For high inflow rate tubes dilated close to the inflow nodes, slowing down flow there and thus increasing absorption. Variance is reduced by two orders of magnitude to $0.040$. (b) Also for the optimal flow rate variance in absorption is reduced to $0.030$. Note, that although metabolite outflux is penalized, it only decreased for (b) to $1.9\%$ and increased for (a) to $1.9\%$ and (c) to $29.0\%$. Note, that no edges were cut.} \label{Fig:Optimized}
\end{minipage}
\end{figure*}
Our toy model is set up to capture the essential flow and transport characteristic along the rows of a two-dimensional network excerpt modulus a geometrical factor. To confirm that the same functional dependence of the optimal inflow rate holds for two-dimensional networks, we return to our simulations of rectangular two-dimensional networks. For a given parameter choice, we vary the inflow rate and determine its optimal value by the minimal variance in absorption. We independently vary the absorption parameter $\gamma$ and the tube radius $R$, equal for all tubes right now, as well as the overall size of the network $L$, see Fig.~\ref{FIG:Leaky}. While one parameter was varied, the other parameters were kept constant.   
To cover a large parameter range, base parameters values are chosen as $\ell=\SI{0.1}{\milli\meter}$, $R=\SI{3}{\micro\meter}$, and $\SI{10}{\per\meter}$, see supplementary material S5.
For each parameter combination the inflow rate was varied in step sizes of $\Delta Q=\SI{1.5e-6}{\milli\meter\cubed\per\second}$. Note, to increase the  overall size of the network additional nodes where added. Therefore, $Q_{\text{out}}$ decreased, and thus the overall flow velocity gradient decreased as well. Each run over a parameter combination was repeated 15 times with different random Gaussian node perturbations of a rectangular two-dimensional network with tubes of the same radii $R$. We find a linear scaling between the optimal inflow rate and the absorption parameter, the radius, and the overall size of the network in agreement with the scaling law's prediction. Even more, if we multiply the optimal inflow rate derived above for the one-dimensional tube network by a geometrical factor $\Gamma$ taking into account the two-dimensional network geometry the numerical results follow exactly the analytical prediction. The geometrical factor is product of three terms $\Gamma=\Gamma_{\text{L}} \cdot \Gamma_{\text{AR}} \cdot \Gamma_{\text{IF}}$, where the first term is correcting the length of the network and the later two are needed to correctly link the velocity profiles with the inflow in the network. The total length of the network in flow direction is effectively shortened as tubes of the two dimensional network are not connected with a angle of $180^{\circ}$ as in the toy model but the network is a tessellation of approximately equilateral triangles. The length of the network has thus to be shortened corresponding to the ratio of height and side length of such a triangle with $\Gamma_{\text{L}}=\sqrt{2}/3$. Considering the rows of nodes, the inflow of fluid in 16 nodes is distributed to 17 nodes in the next row. The flow in tubes connecting these two layers of nodes is thus reduced by the ratio $\Gamma_{\text{AR}}=16/17$. As the total inflow is the inflow over the complete width of the network, the optimal inflow has to flow into every of the 2 to 3 tubes connected to the 16 inflow node giving rise to $\Gamma_{\text{IF}}=47$.

\subsection*{Optimization of network architecture for uniform supply}
We found that a global change in the total fluid inflow rate is the most important control mechanisms to generate uniform supply patterns in a tissue pervaded by a transport network. How does a network architecture need to change to compensate low or high inflow rates? How much more can we minimize the variance in absorption even if fluid flow rate is optimal? To answer these questions we now optimize our previously found supply patterns by allowing for local dilation or contraction of tubes starting with the randomized networks introduced above. In addition to tube dilation and contraction, changes to the network architecture by discarding entire tubes are allowed. A tube is regarded as cut, if it radius is reduced below a threshold of $\SI{0.05}{\micro\meter}$, compared to an average tube radius of R=\SI{3}{\micro\meter}. While locally changing the network architecture, we keep the total amount of material $M=\sum_i R_i \ell_i$ within the network constant as we redistribute changes in $M$ over all radii equally. For this we numerically optimize the network topology using Monte-Carlo methods, explained in detail in supplementary material S3B.  

We optimize the network architecture regarding uniform tissue supply for the cases of low, high and optimal inflow rate, see Fig.~\ref{Fig:Optimized}. In all three cases overall variance in absorption was successfully decreased. For low inflow rates, we observe a contraction of tubes near the influx nodes and an expansion of tubes toward the opposing end. Contraction of tubes speeds up flow velocities thus reducing otherwise dominating absorption close to the inflow nodes and thereby making metabolites available for absorption further onwards. The increase in absorption follows spatially the rapid increase in radius. This indicates that shifts in the radius distribution impact the local absorption profile strongly. 
For high inflow rates, we observe the opposite optimization mechanism. Tubes dilate close to the inflow and contract toward the opposing end. Here, dilation decreases flow rate and increases the absorption early on, while at the same time reducing the amount of metabolite flushing through. For the optimal inflow rate, we observe slight dilation near the inflow as well as near the outflow nodes. These changes correct for network artifacts arising from the chosen rectangular form of the excerpt. In all optimized networks we find small fluctuations in the absorption pattern which result from the randomized node positions and random tube lengths.
\section*{Discussion}
We investigated what is needed to achieve a uniform supply rate of metabolites to tissue via a tubular transport network. We find that the fluid inflow rate is the most important control mechanism. We give an analytical scaling law for the optimal inflow rate as a function of system parameters. Yet, even if the optimal inflow rate is not available, altering the network geometry by dilating or contracting certain tube radii can reduce the overall variance in supply by an order of magnitude.

Optimizing for uniform supply rate across a transport network is a novel perspective regarding the theoretical investigation of optimal transport networks, where the focus is mainly on minimizing total dissipation \linebreak $P=\sum_{\text{i}} Q_{\text{i}}^2/K_{\text{hyd,i}}=\sum_{\text{i}}\pi \langle U\rangle_{\text{i}}^2\eta \ell_{\text{i}}$ \cite{durand_structure_2007,banavar_topology_2000,katifori_damage_2010,corson_fluctuations_2010}. For comparison, we compute the total dissipation for our example network shown in Figs.~\ref{Fig:Uniform}, \ref{Fig:Optimized}. For the networks of equal radii, we find that the dissipation for the optimal inflow rate is of the same order of magnitude to hundredfold higher than for the less uniform supply patterns arising from low and high inflow rates, respectively. Optimizing the network architectures to enhance uniform metabolite supply even increases total dissipation for low and optimal inflow rate, while dissipation is only slightly decreased for high inflow rate. We conclude that total dissipation and uniform metabolite supply are orthogonal properties regarding transport networks. It could well be that biological transport networks balance both properties by optimizing them at the same time. Yet, we observe the differentiation of biological transport networks into different types of tubes like lower order versus higher order veins. This suggest that biological transport networks could be divided into parts that are targeted at transport costs, others targeted at mechanical structures, and others targeted at supply.

We find that inflow rate into a tissue has the biggest impact on how uniform supply is throughout the tissue. For plants, sub-optimal environments, such as a drought, lead to a reduction of water flow. Following our results, a change in the inflow rate will result in a change in the supply pattern, even if the same amount of metabolite is still available. Plant leaves can actively control fluid flow rates by managing evaporation via opening and closing of stomata. 
It is inspiring to note that therefore plants could control to some extend for optimal inflow rates. Unfortunately, to our knowledge, no data on flow rates in leaf veins is available to test this. Alternatively, we find that specific patterns of vein radii could also compensate sub-optimal inflow. Though the adaptation of xylem veins on drought conditions has received general attention, see e.g. \cite{alvarez_metabolomic_2008, EilmannDrought, cannyXylemCons, lovisolo_effects_1998}, it has not been investigated to what extend plants modify the hierarchy of higher order xylem vessel radii for compensatory patterns when grown in sub-optimal conditions. Although our findings predict the regulation of the flow rate by stomata control to be the dominant mechanism, it would be fascinating to check for radii patterns in higher order vessels with established means of vessel network analysis \cite{ronellenfitsch_topological_2015}.

We investigated uniform metabolite supply by xylem vein vasculature focusing on two dimensions. On the level of modeling metabolite absorption on vessel walls, our framework can readily be extended to network topologies embedded in a three dimensional space. That said, the dynamics of metabolite supply within the tissue surrounding the vessel walls changes dramatically if we go from two to three dimensions, simply because the physical space to be supplied increases. The spatial distribution of metabolite concentration in the tissue can be resolved by an explicit treatment of the reaction-diffusion dynamics in the extravascular space. Here, for example the Krogh formalism allows to reduce this computationally complex task to the spacing between vessels as an additional parameter \cite{krogh_number_1919,Rubenstein:2012}. To our knowledge the concept of Krogh radii has not been considered in plant tissue, yet. As we consider flat leaves, we restricted our analysis here to two dimensions. We studied vascular networks with biological observed vessel spacing of $\langle \ell \rangle\approx \SI{0.1}{\milli\meter}$ in our model, assuming that metabolite spread for these physiological values is not limited within the tissue but rather limited by the supply through the vasculature. Since we investigate uniform supply patterns the variation in absorption rates of neighboring vessels is by definition very small, also limiting  supply variations in the tissue. 

Since leaf vascular specifics have been incorporated in our model using a distinct source and sink distribution on the network level, the derived scaling law is only applicable to higher order xylem vessel networks. However, the chosen hydrodynamic perspective of the metabolite spread through a vascular system considers only few assumption and thus allows for a board applicability also in other biological systems. As such the absorption along a tube can be discussed in the setting of capillary beds in animal vasculature. Here, metabolites may be actively transported across the vessel wall with potentially non-linear reaction kinetics that we in this work only approximate by a linear absorption parameter. More importantly, vessels are so small, that blood flows in a plug flow and not Poiseuille flow. In our theoretical work, Poiseuille flow is the key to generate the fast mixing of metabolites across a tube, which is impaired in pure plug flow. However, blood cells being squeezed through the tiny vessel create turbulent eddies and recirculation zones in the flow, which drive fast mixing across a vessel \cite{Rubenstein:2012}. Based on fast mixing, our results may very well be applicable to capillary beds. For capillary beds, inflow rates are to first approximation a function of heart rate and the allocation of fluid along the hierarchical circulatory system. However, capillary beds also auto-regulate their flow by dilating or contracting so-called sphincters situated at the inflow nodes that dilate or contract the close by capillaries \cite{Rubenstein:2012} - a control mechanism in agreement with our findings. Note, that in this example even the location of compensatory regulation close to the inflow follows the predictions of our hydrodynamic model. 

Taken together the evidence of control mechanisms in plant and animal vasculature, albeit scarce, suggests that indeed uniform supply might very well be targeted at the level of higher order veins and capillary beds. Our scaling law predicts a simple relationship between inflow rate and tissue size or vessel radius. Thereby, we pave the way to experimentally investigate supply patterns in biological transport networks. 

Transport networks are at the basis of not only biological organisms but also technological design and medical application. Investigating what properties make a transport network give rise to uniform supply we identify the most important control mechanism, mainly inflow rate and secondary vessel diameter close to inlets. These controls may be important for so many more transport system than the ones exemplified here. But most importantly it sheds light on our understanding of the \emph{transport} dynamics and not just fluid flow profiles in transport networks.

\section*{Acknowledgements}
We like to thank M.~P.~Brenner and P.~Cha for initial discussions about absorption in single tubes. This research was supported in part by the Deutsche Forschungsgemeinschaft (DFG) via grant SFB-937/A19 and the Max Planck Society.%

\pagebreak
\foreach \x in {1,...,8}
{%
\clearpage
\includepdf[pages={\x,{}}]{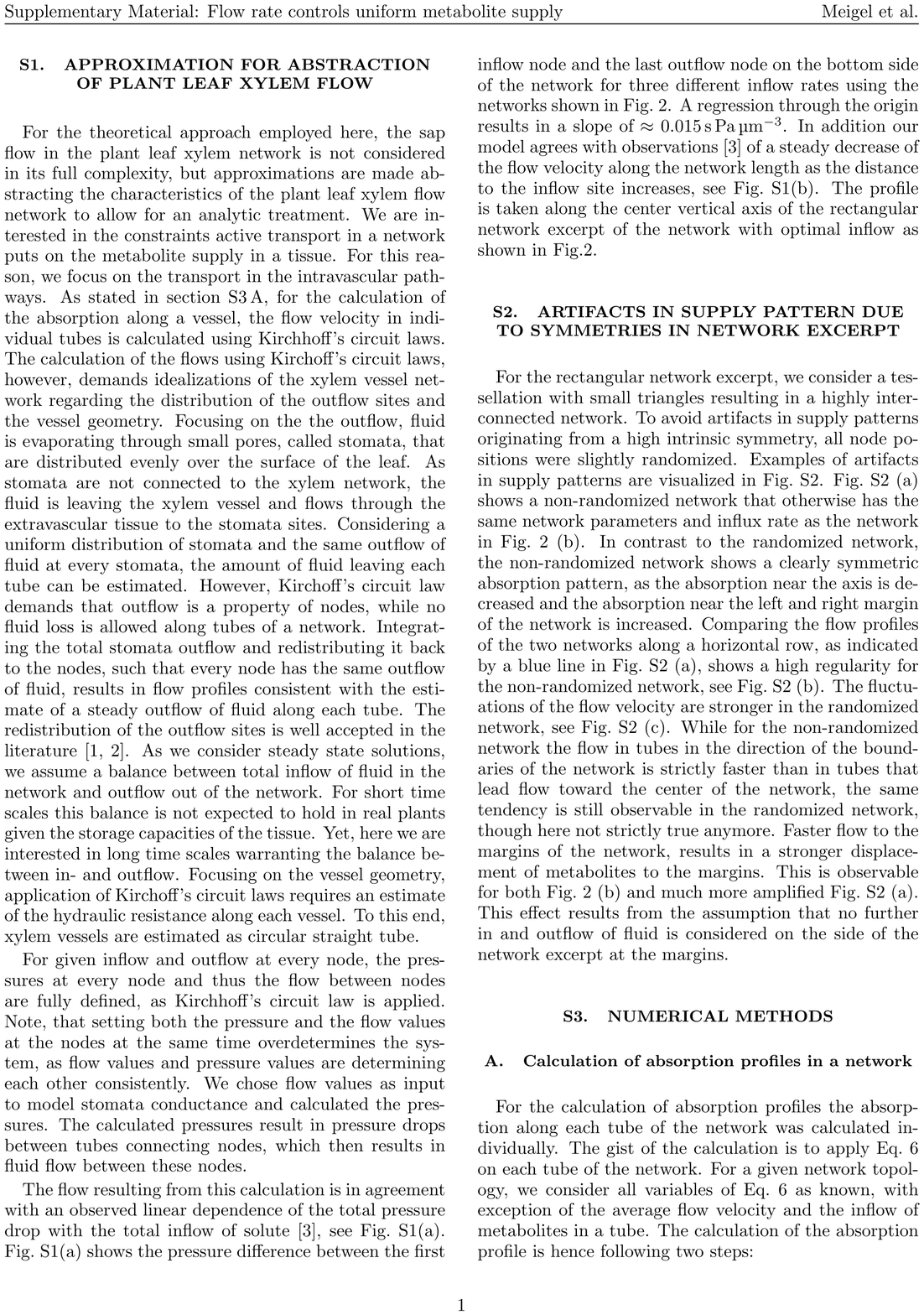}
}

\end{document}